\documentclass[sigplan,screen,version]{acmart}

\title{Incremental Proof Development in Dafny with Module-Based Induction}

\author{Son Ho}
\affiliation{\institution{Inria}\country{France}}

\author{Clément Pit-Claudel}
\affiliation{\institution{EPFL}\country{Switzerland}}

\usepackage{xparse}
\NewDocumentCommand{\li}{v}{\textbf{\footnotesize\texttt{#1}}}

\newcommand\myparagraph[1]{\emph{#1}.}

\usepackage{xspace}
\newcommand{\fstar}{F$^\ast$\xspace}
\newcommand{\comments}{true}
\usepackage{ifthen}
\ifthenelse{\equal{\comments}{true}}
{}

\usepackage[frozencache=true]{minted}

\AtBeginEnvironment{minted}{\fontsize{9}{10}\selectfont}

\begin{document}

\begin{abstract}

Highly automated theorem provers like Dafny allow users to prove simple
properties with little effort, making it easy to quickly sketch proofs.
The drawback is that such provers leave users with little control about the
proof search, meaning that the small changes inherent to the iterative process
of writing a proof often lead to unpredictable variations in verification time,
and eventually hard-to-diagnose proof failures.
This sometimes turns the boon of high automation into a curse, as instead of
breaking early and showing unsolved goals to the user like in Coq, proofs tend
to gradually become unstable until their verification time explodes. At this
point, the absence of a proof context to investigate often leaves the user to a
painful debugging session.
In this paper, we show how to use Dafny modules to encode Coq-like induction
principles to dramatically improve the stability and maintainability of proofs
about inductive data structures.
\end{abstract}

\maketitle

\section{Introduction}

Writing a mechanized proof generally implies working in an incremental manner, by
first laying out a simplified version of the problem under study and shaping
the proofs, then gradually complexifying the definitions while
updating those proofs.
For instance, to verify a compiler one might start with a simplified version of
the source AST that omits the complex cases, before adding those cases one by one.
Highly automated theorem provers like Dafny make the first iterations
pleasant as the prover manages to discharge most of the then-simple proof
obligations.
Unfortunately, as the problem gets more complicated, automated proofs can become
unstable or simply break.
Worse, most automated theorem provers only offer limited tools to debug unstable
or failing proofs: unlike an ITP like Coq, an ATP like Dafny does not show a
goal or a proof context along with each failure.
In an attempt to recover the benefits of ITPs, we demonstrate in this short
paper how to structure inductive proofs in Dafny by using modules to encode
Coq-like induction principles.
We report on our experience of iterating through the proofs of a prototype of a
self-hosted compiler for the Dafny language, and on the benefits in terms of
proof maintainability and stability.

\section{Inductive Proofs in Dafny and Coq}
\label{sec:issues}

\myparagraph{Comparing Dafny and Coq}
Let us start with a minimal example: a proof that list concatenation is
associative, in Dafny and Coq (listing \ref{lst:simple_proofs}).
In Dafny, we define the concatenation as \li+App+ (``append'') and do the proof
by induction in \li+Assoc+.
Dafny automatically discharges the \li+Nil+ case.
In the \li+Cons+ case, we use a recursive call
(line~\ref{AppAssoc:rec}) to invoke the induction
hypothesis.

\begin{figure}
\begin{minipage}[t]{0.45\textwidth}
\centering
\begin{minted}[linenos,mathescape=true,escapeinside=~~,texcomments,numbersep=0pt]{dafny}
  // Dafny
  datatype LL<T> =
  | Nil | Cons(h: T, t: LL<T>)

  function App<T>(
    l0: LL<T>, l1: LL<T>) : LL<T>
  { match l0
    case Nil => l1
    case Cons(h, t) => Cons(h, App(t, l1)) }

  lemma Assoc<T>(l0: LL<T>,
    l1: LL<T>, l2: LL<T>
  ) ensures App(App(l0, l1), l2)
         == App(l0, App(l1, l2))
  { match l0 ~\label{AppAssoc:Match}~
    case Nil => // Nothing to do
    case Cons(h, t) => Assoc(t, l1, l2); } ~\label{AppAssoc:rec}~
\end{minted}
\begin{minted}[mathescape=true,escapeinside=~~]{coq}
  (* Coq *)
  Inductive ll T :=
  | Nil | Cons : T -> ll T -> ll T.

  Fixpoint app {T} (l0 l1: ll T) : ll T :=
    match l0 with
    | Nil => l1
    | Cons h t => Cons h (app t l1)
    end.

  Lemma assoc {T} (l0 l1 l2 : ll T) :
    app (app l0 l1) l2 = app l0 (app l1 l2).
  Proof.
    induction ls0 as [|hd tl IH]; intros; simpl. ~\label{app_assoc:induction}~
    + reflexivity. ~\label{app_assoc:nil}~(* Nil case *)
    + rewrite IH. reflexivity. ~\label{app_assoc:cons}~(*Cons case *)
  Qed.
\end{minted}
\end{minipage}
\captionof{listing}{Proof that list concatenation is associative (top: Dafny,
   bottom: Coq).}
\label{lst:simple_proofs}
\end{figure}

In Coq, we use \li+induction ls0 as ...+ to invoke the induction principle that
Coq automatically derived from the definition of \li+ll+, which, after a call to
\li+simpl+ to simplify the context, gives us two goals:

\begin{math}
\footnotesize
\texttt{T} : \texttt{Type},\, ls_1: \texttt{ll}\; \texttt{T},\, ls_2 : \texttt{ll}\; \texttt{T} \vdash
\texttt{app}\; ls_1\; ls_2 = \texttt{app}\; ls_1\; ls_2
\end{math}

for the \li+Nil+ case, and:

\begin{math}
\footnotesize
\dots,\;
\texttt{IH} : \forall ls_1\; ls_2.\; \texttt{app}\; (\texttt{app}\; t\; ls_1)\; ls_2 =
\texttt{app}\; t\; (\texttt{app}\; ls_1\; ls_2)\\
\indent\vdash
\texttt{Cons}\; h\; (\texttt{app}\; (\texttt{app}\; t\; ls_1)\; ls_2) =
\texttt{Cons}\; h\; (\texttt{app}\; t\; (\texttt{app}\; ls_1\; ls_2))
\end{math}

for the \li+Cons+ case.

From there, it is easy to determine how to invoke the induction hypothesis
(\li+rewrite IH+, with unification filling in the arguments) in the \li+Cons+
case.
We finally conclude both goals by reflexivity.
In contrast, in Dafny: 1. we have to write the inductive structure by hand; 2. we
are not shown goals; 3. we cannot use unification to instantiate the induction
hypothesis, and must instead specify arguments to the recursive call.
This isn't an issue when doing simple proofs, but is a significant
burden when working on more realistic cases.
In particular, Dafny doesn't provide much information to the user when a proof
breaks, while Coq displays the precise goal on which it got stuck.
As a result, the user often spends a significant amount of time debugging
broken proofs to understand \emph{which} proof obligation failed, before actually
spending time on fixing it.

Another issue is proof duplication. We notice that in practice most proofs about
lists follow the same structure: we perform an induction,
and in the \li+Cons+
case call the induction hypothesis on the tail of the list; writing the
inductive structure and specifying the arguments to the recursive call by hand
leads to a lot of boilerplate in Dafny.

One final issue is proof evolution.  In Coq, adding an additional constructor,
maybe
\begin{math}
\footnotesize
\texttt{Snoc}: \texttt{ll}\; \texttt{T} \rightarrow \texttt{T} \rightarrow
\texttt{ll}\; \texttt{T}
\end{math},
would lead to a failed proof with a new
unsolved goal clearly identifying what is missing.
Dafny, in contrast, would
first try to derive a contradiction in the missing case (\li+Snoc+), and having
failed to do that would report the missing case (without showing a proof
context).
This contradiction proof can be costly: in extreme cases, e.g. with
many other constructors, it can fail to complete and we may simply get an
unspecific proof failure for the whole lemma, without further details.
A solution might be to forbid the users from omitting cases in a match; this is however
not desirable in practice, as programmers often use this convenience to avoid considering
many irrelevant cases in their proofs.
\\

\myparagraph{Using an induction principle in Dafny}
We propose to structure the proofs with an induction
principle, by which we factor out the structure of the proofs and decompose the
various proof obligations, leading to less mundane work, a better debugging
experience and finer control over verification times.
The use of induction principles is inspired by provers like Coq which generate
them for free; in the case of Dafny we have to write them by hand.
\footnote{The induction principle we introduce here is actually simplistic for the
purpose of clarity. For more realistic versions, see Section~\ref{sec:mini-dafny}.}

\begin{figure}
\begin{minipage}[t]{0.5\textwidth}
\centering
\begin{minted}{dafny}
abstract module ListInduction {
  predicate P<T>(ls: LL<T>)

  lemma Induct_Nil<T>()
    ensures P<T>(Nil)

  lemma Induct_Cons<T>(
    h: T, t: LL<T>
  ) requires P(t)
    ensures P(Cons(h, t))

  lemma Induct<T>(ls: LL<T>)
    ensures P(ls) {
  match ls
  case Nil => Induct_Nil<T>();
  case Cons(h, t) => Induct(t); Induct_Cons(h, t); } }

module AppAssoc refines ListInduction {
  predicate P ...
  { forall l1, l2 ::
      App(App(ls, l1), l2) ==
      App(ls, App(l1, l2)) }

  // "..." is Dafny syntax to reuse the
  // signatures defined in the module
  // ListInduction
  lemma Induct_Nil ... {}
  lemma Induct_Cons ... {} }
\end{minted}
 \end{minipage}
 \captionof{listing}{Proof of associativity of list concatenation with a module-based
   induction principle.}
\label{lst:ind_princ}
\end{figure}

We proceed by definining an induction principle for lists in the form of an
abstract module (\li+ListInduction+ in listing~\ref{lst:ind_princ}).
We declare the target property that we wish to prove by induction through an
abstract predicate \li+P+, together with the rules it must
satisfy: \li+Induct_Nil+ states that \li+P+ must be true for the empty list, and
\li+Induct_Cons+ states that it must be true on non-empty lists provided it its
true on their tails.
Those abstract declarations act like holes: the corresponding proofs are to be filled later.
Given those assumptions, we can prove by induction once and for all that
\li+P+ always holds (lemma \li+Induct+).

This abstract module provides a generic structure for all the inductive proofs
for lists; in particular we can use it to prove associativity.
We do so by defining a module named \li+AppAssoc+ which refines
\li+ListInduction+.
This time, we have to fill in the blanks: we state the associativity property by
providing a definition for \li+P+, and write proofs for \li+Induct_Nil+ and
\li+Induct_Cons+; as Dafny manages to discharge the proofs automatically, they
are empty. Note that because the proofs are empty, Dafny actually allows us to omit those
lemmas, which is in practice very useful when there are a lot of trivial cases; Dafny
would however report an error if it fails to prove on its own a lemma that we omitted.
The theorem we want is finally given by \li+AppAssoc.Induct+, that we can use
without additional work.
Using an induction principle for this simple example might seem overkill; we
illustrate the benefits on more realistic examples in the next sections.

\section{Applying the Induction Principle on Mini-Dafny}
\label{sec:mini-dafny}

\subsection{Verifying IsPure}

We explained how to define and use an induction principle on the simplistic example of
list concatenation. Let us now illustrate how it can be adapted to a more interesting
example, namely verifying micro-passes of a compiler for a simple language based on Dafny
that we call mini-Dafny. We make the whole development available in the companion artifact~\cite{artifact}.
We adapted this language from a work-in-progress verified compiler for the Dafny
programming language~\cite{compiler-bootstrap}. The problem of verifying mutiple
compilation passes, which involved repeatedly proving inductive theorems involving the
same, big function (\li+InterpStmt+ is around 1000 LoCs in~\cite{compiler-bootstrap}) provided
the initial motivation for the present work.

\begin{minted}[linenos,mathescape=true,escapeinside=||,numbersep=-5pt]{dafny}
  function InterpStmt(s: Stmt, ctx: Context):
    Result<(int, Context)> {
    match s {
    case Bind(bvar, bval, body) =>
      // ':-' below is a monadic bind
      var (bvalv, ctx1) :- InterpStmt(bval, ctx);  |\label{InterpStmt:Value}|
      var ctx2 := ctx1[bvar := bvalv]; |\label{InterpStmt:Augment}|
      var (bodyv, ctx3) :- InterpStmt(body, ctx2); |\label{InterpStmt:Body}|
      var ctx4 := ctx1 + (ctx3 - {bvar}); |\label{InterpStmt:Reset}|
      Success((bodyv, ctx4))

    case Seq(s1, s2) =>
      var (_, ctx1) :- InterpStmt(s1, ctx); |\label{InterpStmt:Seq1}|
      InterpStmt(s2, ctx1) |\label{InterpStmt:Seq2}|

    ... /* Omitted */ } }
\end{minted}

We define the semantics of mini-Dafny with an interpreter (\li+InterpStmt+).
For simplicity, values are integers, and we omit all side effects
but in-place updates to local variables.
\li+InterpStmt+ takes as inputs a statement and a context, which is a map from
variable names to integer values, and returns the result of evaluating the
statement together with an updated context.
In order to evaluate a variable declaration (\li+Bind+ case), we first evaluate the bound
value \li+bval+ (line~\ref{InterpStmt:Value}), where
\li+:-+ is a bind for the error monad, meaning the statement \li+var x :- y; st+ is
desugared to \li+match y { case Fail e => Fail e; case Return(x) => st; }+.
We then augment the context with a new binding
for \li+bvar+ (line~\ref{InterpStmt:Augment}),
evaluate the body in this new context (line~\ref{InterpStmt:Body}), and finally
reset the value bound to \li+bvar+ (if this binding exists in the initial
context), so that the bound variable doesn't escape its scope
(line~\ref{InterpStmt:Reset}).
Specifically, at line~\ref{InterpStmt:Reset}, \li+ctx3 - {bvar}+ is the map \li+ctx3+ where we
remove the binding for \li+bvar+ (if it exists), and \li{ctx1 + (ctx3 - {bvar})} is \li+ctx1+
extended with the bindings from \li+ctx3 - {bvar}+ (if a binding exists in both maps, we
take the one from \li+ctx3 - {bvar}+).
Finally, evaluating a sequence of statements (\li+Seq+ case) simply requires chaining
contexts between the statements of the sequence.

Given those semantics for mini-Dafny, we can verify a first micro-pass which
rewrites statements of the form \li+0 * s+ or \li+s * 0+ to \li+0+, provided
\li+s+ is pure (i.e., it doesn't update any local variable);
note that in mini-Dafny we mix statements and expressions.
We first define a predicate \li+IsPure(s: Stmt, locals: set<string>)+ which
states that statement \li+s+ doesn't update variables but the ones listed in
\li+locals+; we use \li+locals+ to track variables bound in declarations and
whose updates won't escape their scope.
In particular, if \li+IsPure(s, {})+ is true then \li+s+ doesn't have side
effects.
Looking at the definition, a declaration (\li+Bind+) is pure if the bound statement
\li+bval+ only updates variables from \li+locals+, and if its body
only updates variables from the set \li{{bvar} + locals}.
An in-place update (\li+Assign+) is pure if it updates the value of a
variable from \li+locals+.
A sequence is pure if it is made of pure statements. We omit the other,
straightforward cases.
For instance, \li+x := 3+ is not pure, while
\li+var x := 0; x := 3+ is pure because \li+x+ is locally bound and won't escape
its scope.

\begin{minted}{dafny}
predicate IsPure(
  s: Stmt, locals: set<string> := {}) {
  match s
  case Bind(bvar: Var, bval: Stmt, body: Stmt) =>
    IsPure(bval, locals) && IsPure(body, {bvar} + locals)
  case Assign(avar, aval) =>
    avar in locals && IsPure(aval, locals)
  case Seq(s1, s2) =>
    IsPure(s1, locals) && IsPure(s2, locals)
  ... /* Omitted */ }
\end{minted}

Now suppose we want to prove the correctness of \li+IsPure+, meaning that
if a statement is pure in the sense of \li+IsPure+ then
evaluating it leaves the context unchanged.
As the proof proceeds by induction over mini-Dafny statements
\footnote{mini-Dafny doesn't have loops, which would require induction over
semantic derivations.} we introduce an induction principle to
reason over the mini-Dafny AST.

\begin{minted}{dafny}
predicate P(st: S, s: Stmt)
predicate P_Step(st: S, s: Stmt, st1: S, v: V)
... // Some definitions omitted

lemma P_Step_Sound(st: S, s: Stmt, st1: S, v: V)
  requires P_Step(st, s, st1, v)
  ensures P(st, s)

lemma InductSeq_Step(
  st: S, s: Stmt, s1: Stmt, s2: Stmt, st1: S, v1: V)
  requires s == Seq(s1, s2)
  requires P_Step(st, s1, st1, v1)
  requires P(st1, s2)
  ensures P(st, s)

... // Omitted: lemmas for the various inductive cases
\end{minted}

Defining an induction principle for mini-Dafny requires a bit more work than for lists.
We require an abstract \li+P(st: S, s: Stmt)+ predicate which states the target
property for statement \li+s+ in state \li+st+.
Importantly, we use an abstract type \li+S+ for the states, because the
user might want to carry more information than just a single context.
We do the same for values, for similar reasons.
We also require an auxiliary predicate \li+P_Step+ to mention intermediary
steps of execution.
The \li+P_Step(ctx: S, s: Stmt, ctx1: S, v: int)+ predicate states that evaluating \li+s+
starting in state \li+ctx+ succeeds and yields a new state \li+ctx1+ and a value
\li+v+;
as we need to link it to \li+P+ somehow, we also require
that \li+P_Step+ implies \li+P+ through the (abstract) lemma \li+P_Step_Sound+.
We then decompose the inductive cases into precise lemmas.
For instance, \li+InductSeq_Step+ states that, if the target property holds
for the execution of \li+s1+ starting in \li+st+, and also holds for the execution
of \li+s2+ starting in the state resulting from executing \li+s1+, then it holds for the
whole sequence \li+s1; s2+, starting in \li+st+.
We finally show how to use this induction principle for the proof of correctness
of \li+IsPure+.

\smallskip
\begin{minipage}[t]{0.48\textwidth}
\centering
\begin{minted}[linenos,mathescape=true,escapeinside=||,numbersep=5pt]{dafny}
datatype S =
  S(locals: set<string>, ctx: Context)
type V = int

predicate SameCtxs(locals, ctx, ctx1) {
  && ctx1.Keys == ctx.Keys
  && ctx1 - locals == ctx - locals }

predicate P_Step(st, s, st1, v) {
  && IsPure(s, st.locals)
  && st.ctx.Keys >= st.locals
  && InterpStmt(s, st.ctx) == Success((v, st1.ctx))|\label{IsPureProof:interpEq}|
  && st1.locals == st.locals
  && SameCtxs(st.locals, st.ctx, st1.ctx) }

predicate P(st, s) {
  IsPure(s, st.locals) ==> |\label{IsPureProof:pure}|
  st.ctx.Keys >= st.locals ==> |\label{IsPureProof:locals}|
  match InterpStmt(s, st.ctx) {
    case Failure _ => true
    case Success((_, ctx1)) => |\label{IsPureProof:success}|
      SameCtxs(st.locals, st.ctx, ctx1) |\label{IsPureProof:same}|
  } }
\end{minted}
\end{minipage}
\bigskip

We define the state as a pair of a set of variable names and a context.
The predicate \li+P+ states that if \li+s+ only updates
variables from \li+st.locals+ according to \li+IsPure+ (line~\ref{IsPureProof:pure}),
and if the context has bindings for the variables listed in \li+st.locals+
(line~\ref{IsPureProof:locals}), then evaluating \li+s+ starting in \li+st.ctx+
yields (if it succeeds) a context which is unchanged but on the variables listed in
\li+st.locals+ (line~\ref{IsPureProof:same}).
We need the condition \li+st.ctx.Keys >= st.locals+ to
ensure that \li+InterpStmt+ won't fail while accessing a variable listed in
\li+st.locals+ because it is undefined.
The predicate \li+P_Step+ is similar to \li+P+ except for the condition that executing
\li+s+ in \li+st+ must succeed, yielding \li+v+ \li+st1+
(line~\ref{IsPureProof:interpEq}), and for the conjunctions which replace
implications (this is slightly technical: suffices to say that \li+P_Step+ must
\emph{unconditionally} state that the execution succeeds; we omit the rule which
enforces this).
Overall, instantiating the induction principle for \li+IsPure+ is
straightforward, and all the proofs go through
automatically.

\subsection{Experience Using the Induction Principle}

We now report on our experience of using the induction principle in practice.
We applied the induction principle to several proofs, namely:
1. \li+IsPure+;
2. \li+EliminateMulZero+: a micro-pass which simplifies statements of the
  form \li+0 * s+ or \li+s * 0+ to \li+0+, provided \li+s+ is pure;
3. \li+UnchangedVar+: a predicate which states that a specific variable is left
  unchanged by an statement.

We made several changes to the mini-Dafny AST in order to evaluate the
cost of updating the proofs:
1. we updated \li+Seq+ to contain an arbitrary number of statements
  ($\footnotesize\texttt{Seq(Stmt, Stmt)} \rightarrow \texttt{Seq(seq<Stmt>)}$).
2. we updated \li+Bind+ (and \li+Assign+) to allow multiple declarations
(assignments, respectively) at once.

We initially introduced \li+UnchangedVar+ to reason about variable inlining
when working on the more mature version of the
compiler~\cite{compiler-bootstrap}.
Though simple in appearance, this property is actually subtle and led to
expensive proofs; we thus resorted to using a module-based induction
principle.
In practice, this approach allowed us to dramatically decrease the time we
spent on maintaining the proofs.

\myparagraph{Factoring out proofs} By using an induction principle we don't have to write
the inductive structure of the proofs by hand, and even get automatic variable
introduction; after we paid the cost of writing this principle, we thus recover similar
advantages to using a tactic like \li+induction+ in Coq.
Instantiating the induction principle by providing definitions for the abstract
declarations (e.g., \li+P+) requires some boilerplate, especially as we
introduced more abstract definitions with each language extension.
In practice, however, this work was straightforward, especially as mistakes done when
instantiating the induction principle were easy to debug and fix, and in particular
easier to fix than the version of the proofs which did not use an induction principle.
We also noticed that our instantiations shared similarities: we might leverage
this fact to reduce the work even further in the future.

\myparagraph{Easy debugging}
Because we wrote the rules required by the induction principle so that they are
small and precise, we were able to quickly pinpoint the reasons behind a failure
whenever a proof broke.
In particular, Dafny would tell us which specific lemma (and thus inductive case) failed.
This allowed us to easily fix proofs when updating the language, and proved
useful when sketching the proofs in the first place, as we could quickly iterate
by adjusting the way we stated the properties we targeted to prove until we got
them right.

\myparagraph{Smooth iterations}
Updating the mini-Dafny language required us to update the induction principle
several times, either by modifying specific rule statements, or by adding more
rules and abstract definitions.
As a result, we could focus on specific changes while updating the proofs, and
the fact that the rules are small and simple made them more stable.
In practice, updating the mini-Dafny language only required us to provide
definitions for the new declarations we introduced in the induction principle,
and to add \emph{one} assertion at one location in the proof of
\li+VarUnchanged+, to guide the SMT solver in its search.

\section{Related Work}

Previous work explored the design of proofs robust to
changes, typically by introducing abstractions and interfaces
\cite{sel4-2014, raft-2016, coq-modules, coq-typeclasses}.
Some work explored the problem of automating inductive proofs altogether by means of
heuristics~\cite{dafny-ind, boyer_moore, acl2, rippling-review, rippling-case, zeno};
in our case, we target properties which are usually too complex to be fully automated.
Other work explored the problem of writing usable inversion theorems but for
ITPs like Coq, for instance to control the size of the generated proof
terms~\cite{monin-2010, monin-2013}.
An obvious way of overcoming the limitations described in Section~\ref{sec:issues} is to
extend ATPs with tactics, as done in~\cite{meta-fstar, why3-tactics, why3-refl}.
In the context of the present work extending the Dafny prover was however not an
option, but it would be interesting to investigate how our approach compares with
using tactics in a tool which supports both (like \fstar or Why3).
Interestingly, if the use of heavy automation has been promoted a lot to
stabilize proofs when using
ITPs~\cite{CPDT, sledgehammer, sledgehammerExp, HammerForCoq, flyspeck}, much
less work went into the problem of stabilizing proofs which already relied on
a high level of automation.
We can however mention attempts to stabilize the proof search itself~\cite{DafnyTriggers}
or preserve VC transformations and proof attempts~\cite{why3-stable}.
Related to the problem of proof maintenance, some recent work explored the
problem of proof repair~\cite{pumpkin-patch, pumpkin-pi}, but in the context of
tactic-based proof assistants like Coq and not in highly automated theorem
provers like Dafny.
Finally, it is worth noting that other works tackled the problem of studying language
semantics in an ATP~\cite{why3-shell, why3-sem, why3-symb}. We however note that none of
those targeted the verification of a multi-pass compiler for a realistic language, which
on our side provided the original motivation for introducing our encodings of induction
principles~\cite{compiler-bootstrap}.

\section{Conclusion}

We demonstrated how to encode an induction principle in Dafny by means of an
abstract module.
By applying this induction principle to case studies taken from iterations over
the proofs of correctness of a compiler for the mini-Dafny language, we showed
that our technique has clear benefits to help the user factor out and maintain
proofs.
In effect, by structuring inductive proofs we relieve the user from
mundane work spent on structuring those proofs, and allow them to focus instead
of their core.
By decomposing inductive proof obligations into small and precise lemmas, we
also make it easy to pinpoint the reason behind proof failures.
As future work, we are planning to investigate how to automate the
process of generating induction principles, as is done in ITPs like Coq.

\bibliography{paper.bib}

\end{document}